\documentclass[twocolumn,times,twocolappendix]{aastex631}

\usepackage{newtxmath}
\usepackage{amsmath}
\newcommand{\msun}{M$_\odot$}
\newcommand{\msunmath}{\text{M}_\odot}
\newcommand{\pr}[1]{\text{Pr}\left(#1\right)}

\DeclareRobustCommand{\VAN}[3]{#2}
\let\VANthebibliography\thebibliography
\def\thebibliography{\DeclareRobustCommand{\VAN}[3]{##3}\VANthebibliography}

\usepackage{enumerate}

\shorttitle{A Deficit of Massive WDs}
\shortauthors{Hallakoun et al.}
\graphicspath{{./}{figures/}}

\begin{document}

\title{A Deficit of Massive White Dwarfs in Gaia Astrometric Binaries}

\correspondingauthor{Na'ama Hallakoun}
\email{naama.hallakoun@weizmann.ac.il}

\author[0000-0002-0430-7793]{Na'ama Hallakoun}
\affiliation{Department of particle physics and astrophysics, Weizmann Institute of Science, Rehovot 7610001, Israel}

\author[0000-0001-9298-8068]{Sahar Shahaf}
\affiliation{Department of particle physics and astrophysics, Weizmann Institute of Science, Rehovot 7610001, Israel}

\author[0000-0002-3569-3391]{Tsevi Mazeh}
\affiliation{School of Physics and Astronomy, Tel-Aviv University, Tel-Aviv 6997801, Israel}

\author[0000-0002-2998-7940]{Silvia Toonen}
\affiliation{Anton Pannekoek Institute for Astronomy, University of Amsterdam, 1090 GE Amsterdam, The Netherlands}

\author[0000-0001-6760-3074]{Sagi Ben-Ami}
\affiliation{Department of particle physics and astrophysics, Weizmann Institute of Science, Rehovot 7610001, Israel}



\begin{abstract}
The third data release of Gaia introduced a large catalog of astrometric binaries, out of which about 3,200 are likely main-sequence stars with a white-dwarf (WD) companion. These binaries are typically found with orbital separations of $\sim 1$\,AU, a separation range that was largely unexplored due to observational challenges. Such systems are likely to have undergone a phase of stable mass transfer while the WD progenitor was on the asymptotic giant branch. Here we study the WD mass distribution of a volume-complete sample of binaries with K/M-dwarf primaries and orbital separations $\sim 1$\,AU. We find that the number of massive WDs relative to the total number of WDs in these systems is smaller by an order of magnitude compared to their occurrence among single WDs in the field. One possible reason can be an implicit selection of the WD mass range if these are indeed post-stable-mass-transfer systems. Another reason can be the lack of merger products in our sample compared to the field, due to the relatively tight orbital separations of these systems. In addition, we find that about $14\%$ of these systems have distant tertiary companions within 1\,pc.
\end{abstract}



\section{Introduction} \label{sec:intro}
Although the vast majority \citep[$\gtrsim 97\,\%$,][]{Althaus_2010} of all stars end their lives as white dwarfs (WDs), the details of some of the most basic properties of their population are still far from consensus. Perhaps the most basic of all is the WD mass distribution, which bears the imprints of the initial mass function, the star formation history, and the stellar evolution (and in particular the initial-to-final mass relation)---of both single and multiple stars.

The most prevalent technique for assessing the masses of WDs involves the fitting of their observed spectroscopic \citep[e.g.][]{Kepler_2015, Kepler_2016, Kepler_2019} and/or photometric \citep[e.g.][]{Kilic_2020, JimenezEsteban_2023} data with theoretical atmospheric models. This fitting process yields the effective temperature and surface gravity of the WD, which can then be converted to mass using theoretical mass-radius relations, or using the estimated radius if the parallax is known. While highly useful and applicable to many WDs, this approach is reliant on theoretical models and, as a result, may be susceptible to systematic errors stemming from potential inaccuracies within these models, such as the exact chemical composition of the WD atmosphere \citep[e.g.][]{Bergeron_2019}. It can also cause problems with low-resolution spectroscopy in the presence of strong magnetic fields \citep[that are more common among massive WDs, e.g.][]{GarciaBerro_2016}, due to absorption-line broadening by Zeeman splitting. Spectroscopic fitting is also problematic in the case of DC-type WDs, that show no absorption lines \citep[e.g.][]{Caron_2023}.

Other techniques for measuring WD masses that do not rely on atmospheric models, such as pulsational-mode analysis \citep[e.g.][]{Corsico_2019} and measuring the gravitational redshift of absorption lines \citep[e.g.][]{Falcon_2010, Falcon_2012}, are only applicable to a small number of WDs. Model-independent masses can also be obtained for WDs in astrometric binary systems \citep[e.g.][]{Gatewood_1978}. Such systems were scarce until very recently, as will be discussed below.

In the absence of nuclear fusion, WDs slowly cool over billions of years. The more massive WDs become cooler and fainter sooner after star formation due to their much shorter progenitor main-sequence (MS) lifetimes (if they are formed through single-star evolution), and earlier onset of core crystallization, which significantly decreases their specific heat and accelerates their cooling rate \citep{Fontaine_2001}. Cool WDs usually have faint absorption lines (if any), which are crucial for estimating their mass using atmospheric models. In addition, their smaller sizes make them intrinsically fainter than their lower-mass counterparts. For these reasons, magnitude-limited samples of WDs are biased against the more massive WDs \citep[$\gtrsim 0.8$\,\msun, e.g.][]{RebassaMansergas_2015}.

In order to derive the WD mass distribution, it is thus crucial to obtain a volume-limited unbiased census of WDs. With the exception of the nearly-complete 20\,pc WD sample \citep[consisting of about 150 WDs, e.g.][]{Toonen_2017}, the known WD sample in the pre-Gaia era was magnitude-limited rather than volume-limited, and had been dominated by serendipitous discoveries and by targeted-yet-biased searches \citep[such as searching for sources with ultraviolet excess, e.g.][]{Green_1986}. The Sloan Digital Sky Survey \citep[SDSS;][]{York_2000} successfully discovered a few tens of thousand WDs \citep[e.g.][]{Kepler_2021}, but the strong and largely unknown selection biases of the sample limited the possible statistical inferences.

It was not until the second data release (DR2) of the Gaia space mission in 2018 \citep{Gaia_2016, Gaia_2018} that a large, unbiased, sample of WDs has been compiled \citep[][and references therein]{GentileFusillo_2019}. The Gaia catalog \citep[along with subsequent data releases, see][]{Gaia_2021, GentileFusillo_2021} allowed the construction of larger volume-limited samples, including the 20\,pc \citep{Hollands_2018} and 40\,pc \citep{Tremblay_2020, McCleery_2020, OBrien_2023} samples, that are nearly complete for single WDs, and the 100\,pc sample, that is considered to be nearly complete spectroscopically for single WDs down to an effective temperature of $\sim 6,000$\,K \citep[i.e.~still a magnitude-limited sample, see][]{JimenezEsteban_2018, Kilic_2020, JimenezEsteban_2023}.

The third data release (DR3) of Gaia \citep{Gaia_2023} was the first to include a catalog of unresolved sources with non-single star solutions \citep{Gaia_Arenou_2023}. Recently, \citet{Shahaf_2024} have identified nearly 3,200 systems that are likely MS+WD binaries. This unprecedentedly large sample of WDs in solved astrometric binaries offers a unique opportunity to derive a model-independent mass distribution of a volume-complete WD sample. In this paper, we use this catalog to define volume-complete samples of MS+WD binaries with measured dynamical masses (Section~\ref{sec:sample}). In Section~\ref{sec:results} we compare the WD mass distribution of these samples with published mass distributions of field WDs within 100\,pc, showing a stark deficit of massive white dwarfs in our astrometric samples. Finally, in Section~\ref{sec:discussion} we discuss several possible explanations for the observed difference between the distributions.

\section{The Sample}\label{sec:sample}

\subsection{The Astrometric Mass-Ratio Function} \label{sec:AMRF}

In this paper we use the Gaia DR3 catalog of unresolved sources with non-single star solutions \citep{Gaia_Arenou_2023}. Since Gaia only detects the motion of the photocenter of the unresolved system, the nature of the companion(s) remains uncertain when it is based on the astrometric solution alone. \citet{Shahaf_2019} introduced a triage method to rule out MS or binary-MS companions by defining a unitless observational parameter, named the `Astrometric Mass-Ratio Function' (AMRF):
\begin{equation}
    \mathcal{A} \equiv \frac{\alpha_0}{\varpi} \bigg(\frac{M_1}{\textrm{M}_\odot}\bigg)^{-1/3} \bigg(\frac{P}{\textrm{yr}}\bigg)^{-2/3}\, , 
    \label{eq:AMRF}
\end{equation}
where $\alpha_0$ is the angular semi-major axis of the photocenter, $\varpi$ is the parallax, $P$ is the orbital period, and $M_1$ is the mass of the primary. This value can be calculated for every astrometric binary with a known $M_1$, e.g.~based on theoretical evolutionary tracks and the location on the Hertzsprung-Russell (HR) diagram.

The measured AMRF values can then be compared to the expected AMRF values for a given combination of mass ratio, $q=M_2/M_1$, and luminosity ratio, $\mathcal{S}=F_2/F_1$, where $F_i$ is the flux of the $i$th component:
\begin{equation}\label{eq:expectedAMRF}
    \mathcal{A} = \frac{q}{(1+q)^{2/3}} \left( 1 - \frac{\mathcal{S}(1+q)}{q(1+\mathcal{S})} \right).
\end{equation}
For a non-luminous companion $\mathcal{S}=0$, and its mass can be extracted from Equation~(\ref{eq:expectedAMRF}) using the measured AMRF value and the estimated $M_1$.

\begin{figure}
    \centering
    \includegraphics[width=\columnwidth]{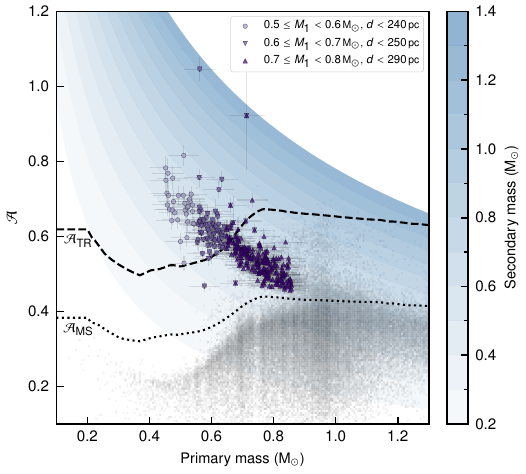}
    \caption{The Gaia DR3 astrometric mass-ratio function (AMRF; $\mathcal{A}$) as a function of the primary mass. The maximal possible AMRF values for an MS secondary ($\mathcal{A}_\text{MS}$) and an MS close-binary ($\mathcal{A}_\text{TR}$) companions are marked by the dotted and dashed lines, respectively.
    The underlying light-gray density plot shows the measured values from Gaia DR3 \citep[after applying some quality cuts on the astrometric solutions, see][]{Shahaf_2023}. Our selected volume-limited samples of high-probability MS+WD systems with different primary masses (see legend) are plotted with purple markers and gray error bars. Note that because of the primary mass uncertainty, there is some overlap between the samples.
    The expected AMRF values of WD companions of different masses are marked by blue-shaded stripes, ranging from 0.2\,\msun\ (light blue) to 1.4\,\msun (dark blue), in steps of 0.1\,\msun.
    }
    \label{fig:AMRF}
\end{figure}

Assuming the most stringent constraints on the mass and luminosity ratios for different companion types, \citet{Shahaf_2023} defined the limiting AMRF values for an MS companion, $\mathcal{A}_\text{MS}$, or an MS+MS binary companion, $\mathcal{A}_\text{TR}$. Figure~\ref{fig:AMRF} shows the full distribution of the measured AMRF values from Gaia DR3 as a function of the primary mass \citep[after applying some quality cuts on the astrometric solutions, see section 3.1 of][]{Shahaf_2023}. The maximal expected AMRF values assuming an MS+MS binary or a hierarchical MS+MS+MS triple system are marked by the dotted and dashed lines, respectively. These values separate the astrometric non-single star sample into three classes: \textit{class I} ($\mathcal{A} < \mathcal{A}_\text{MS}$), where the companion is most likely a single MS star; \textit{class II} ($\mathcal{A}_\text{MS} < \mathcal{A} < \mathcal{A}_\text{TR}$), where the companion cannot be a single MS star, but can still be a close binary of two MS stars; and \textit{class III} ($\mathcal{A}_\text{TR} < \mathcal{A}$), where the companion is neither a single MS star nor a close MS binary, but a likely compact object.

Based on Gaia DR3 data, \citet{Shahaf_2023} complied a sample of high-probability \textit{class-III} binary systems, consisting of an MS and a compact object. Using Gaia synthetic photometry, calculated based on Gaia's low-resolution BP/RP spectra, \citet{Shahaf_2024} identified systems with a significant excess red color to further differentiate between \textit{class-II} hierarchical MS triple systems and binaries with a compact-object companion. While an MS companion(s) is expected to contribute flux in the redder wavelengths, the light contribution of a WD companion to the Gaia band is expected to be negligible (due to the extreme size ratio), unless it is very hot and young, then it contributes to the bluer end of the spectrum \citep[see][for details]{Shahaf_2024}. This procedure resulted in a much more comprehensive catalog of nearly 3,200 high-probability binary systems consisting of an MS and a compact object, most of them in the WD mass range. The orbital separations of these systems, on the order of 1\,AU, place them on a largely unexplored region of parameter space, with only a few such systems known previously \citep{Anguiano_2022, Parsons_2023}. In addition, these systems are not easily reproduced by standard binary population synthesis codes \citep[see discussion in][]{Shahaf_2024}.

Here we focus on a volume-complete sample of likely MS+WD binaries, following the method of \citet[][see Section~\ref{sec:sampleselection} below]{Shahaf_2024}. We should note that this triage method relies on Gaia's astrometric solutions \citep{Halbwachs_2022} along with the primary mass estimates reported by Gaia \citep{Gaia_Arenou_2023}. It is possible that despite the precautions taken by \citet{Shahaf_2023, Shahaf_2024}, some spurious results still got through. However, for the mass ratios discussed in this paper and for MS primaries, the contamination by spurious detections is expected to be small. Spectroscopic follow-up observations of the subsample used here should verify the validity of the results.

\subsection{Sample Selection}\label{sec:sampleselection}
We start by selecting systems from the original catalog of \citet{Shahaf_2023} that are unlikely to contain an MS companion,
\begin{equation}\label{eq:PrI}
    \pr{\text{I}} < 10^{-3},
\end{equation}
where $\pr{\text{I}}$ is the probability of being a \textit{class-I} system. This leaves 7,687 out of 101,380 systems of the cleaned sample of \citet{Shahaf_2023}.

Following \citet{Shahaf_2024}, we select only systems satisfying
\begin{equation}\label{eq:initial_m1max}
    M_1 < 1.2\,\msunmath\ + \Delta M_1,
\end{equation}
where $\Delta M_1$ is the uncertainty of the primary mass estimate. By this we avoid systems with primaries that have already left the MS, rendering their mass estimate based on the location on the Hertzsprung-Russell (HR) diagram unreliable, leaving 6,880 systems.

Next, we remove systems that are likely to be triple-MS systems, by excluding 3,846 probable \textit{class-II} systems with red excess flux (as mentioned above, WD companions are not expected to contribute any detectable flux in the redder wavelengths):
\begin{equation}\label{eq:NCEclassII}
    \left( \pr{\text{II}} \geq 0.5 \right) ~ \& ~ \left(\pr{\text{red}} \geq 0.56\right),
\end{equation}
where $\pr{\text{II}}$ is the probability of being a \textit{class-II} system, and $\pr{\text{red}}$ is the probability for the presence of red excess flux compared to the predicted location of the photometric primary on the HR diagram, assuming a total age of 2\,Gyr and the estimated metallicity of each system \cite[see][for details]{Shahaf_2024}. The red-excess probability was calculated using Gaia's synthetic photometry in the Johnson-Kron-Cousins photometric system over the $V$ vs. $B-I$ plane.
We further exclude additional 145 \textit{class-II} systems that do not have color excess data.
This resulted in a total of 2,889 systems that are likely to contain a WD companion.

To ensure a flux ratio $\lesssim 10\%$ in the Gaia $G$ band between the WD and the MS \citep[see figure~14 of][]{Shahaf_2024}, keeping our dark-companion assumption justified, we focus only on systems with
\begin{equation}
    M_1 \geq 0.5\,\msunmath,
\end{equation}
leaving 2,835 systems. This assumption is crucial for the secondary mass estimates.
On the high-mass end, we limit our focus to systems with
\begin{equation}
    M_1 < 0.8\,\msunmath
\end{equation}
in order to ensure a sensitivity to $\approx 0.6$\,\msun\ companions, and to avoid the artificial excess of systems with $M_1 \approx 1$\,\msun\, resulting from the primary mass estimate process of \citet[][see underlying density plot in Figure~\ref{fig:AMRF}, and discussion in \citealt{Shahaf_2024}]{Gaia_Arenou_2023}. This leaves a total of 1,745 systems.

As discussed in \citet{Shahaf_2024}, this selection procedure based on red-excess probability minimizes the false-positive MS+WD identification rate at the expense of a higher false-negative rate (i.e. some true MS+WD systems are excluded from the sample). Systems with primary (MS) masses smaller than $\sim 0.6$\,\msun\ are more likely to have erroneous metallicity estimates, affecting the red-excess probability calculation \citep{Shahaf_2024}. However, since the expected mass range of WD companions with M-dwarf primaries resides within \textit{class-III} on the AMRF plot (see Figure~\ref{fig:AMRF}), we overcome this missclassification here by applying the red-excess cut on \textit{class-II} systems only (Equation~\ref{eq:NCEclassII}).
True MS+WD systems with primary (MS) masses higher than $\sim 0.6$\,\msun\ are more likely to be excluded due to erroneous primary mass estimates, especially those with masses around $\sim 1$\,\msun\ \citep{Shahaf_2024}. Others can be missed if the primary has started to evolve off the MS (which would affect our red-excess estimation). As mentioned above, we avoid these problems by selecting systems with smaller primary masses.

The Gaia DR3 non-single star catalog contains only sources satisfying \citep[][eq. 16]{Halbwachs_2022}:
\begin{equation}
    \frac{\varpi}{\sigma_\varpi} > \frac{20,000\,\textrm{day}}{P},
\end{equation}
where $\varpi$ and $\sigma_\varpi$ are the parallax and its error, and $P$ is the orbital period.
Translating it using Kepler's law to a constraint on the orbital separation, $a$, we get
\begin{equation}
    \frac{a}{\textrm{AU}} > 0.144 \left( \frac{M_1 + M_2}{\msunmath} \right)^{\frac{1}{3}} \left( \frac{\sigma_\varpi}{\textrm{mas}} \right)^{\frac{2}{3}}
     \left( \frac{d}{\textrm{pc}} \right)^{\frac{2}{3}},
\end{equation}
where $d$ is the distance, and $M_{1,2}$ are the component masses.
For the median parallax error of our subsample ($\approx 0.03$\,mas, with a typical relative error of about $4\%$), and for the limiting case of $M_1=0.8$\,\msun\ and $M_2=1.4$\,\msun, we get
\begin{equation}
    \frac{a}{\textrm{AU}} > 0.018 \left( \frac{d}{\textrm{pc}} \right)^{\frac{2}{3}},
\end{equation}
or $a \gtrsim 0.79$\,AU at a distance of 290\,pc (see below).
We thus select only systems satisfying
\begin{equation}\label{eq:a_min}
    a > 0.8\,\text{AU},
\end{equation}
resulting in 1,611 systems.

\begin{figure}
    \centering
    \includegraphics[width=\columnwidth]{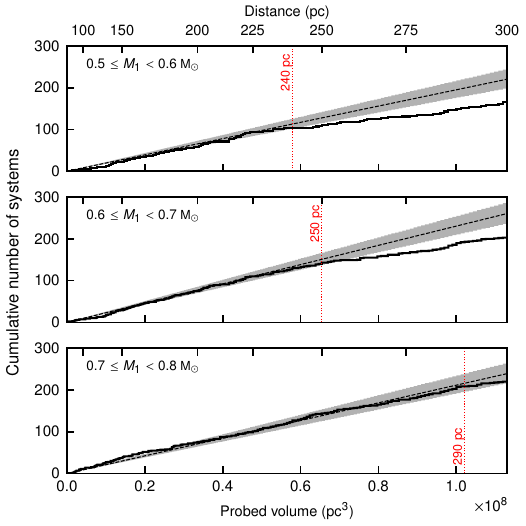}
    \caption{Cumulative number of systems as a function of the probed volume (solid line) for a variety of photometric primary mass bins (\textit{top}: $0.5 \leq M_1 < 0.6$\,\msun, \textit{middle}: $0.6 \leq M_1 < 0.7$\,\msun, \textit{bottom}: $0.7 \leq M_1 < 0.8$\,\msun). The dashed line shows the expected number assuming a constant space density (calculated based on the total number of systems per sample at a maximal distance of 225\,pc), and the gray region highlights its Poisson uncertainty. The dotted red line marks the selected maximal distance for each sample.
    }
    \label{fig:Completeness}
\end{figure}

Next, we evaluate the completeness of our sample.
Figure~\ref{fig:Completeness} shows the cumulative number of systems satisfying the above conditions as a function of the probed volume.
To take into account the Lutz-Kelker bias \citep{Lutz_1973, Luri_2018}, causing the mean observed parallax to be systematically larger than its true value, the distance to each system is taken as the median value of $1/\varpi$ calculated based on 10,000 realizations of the parallax drawn from a Gaussian distribution defined using the observed Gaia parallax value and uncertainty.

For a complete sample, we assume that the total number of systems grows linearly with the enclosed volume, i.e. a uniform number density. We estimate the number density based on the total number of systems within 225\,pc, and check at which distance the cumulative number of systems significantly deviates from the expected number of systems, assuming Poisson uncertainty. We do not fit the cumulative distribution directly, as its steps are not statistically independent \citep[e.g.][]{Hollands_2018}. We check the completeness of the sample for different photometric primary mass bins, and find that the subsample of systems with photometric primaries with $0.5 \leq M_1 < 0.6$\,\msun\ (within $1\sigma$ uncertainty) is more or less complete to a distance of $\approx 240$\,pc, the subsample of systems with photometric primaries of $0.6 \leq M_1 < 0.7$\,\msun\ is complete within a distance of $\approx 250$\,pc, and the subsample of systems with $0.7 \leq M_1 < 0.8$\,\msun\ is complete within a distance of $\approx 290$\,pc.
We thus define three volume-complete subsamples,
\begin{enumerate}[(i)]
    \item $0.5 \leq M_1 < 0.6$\,\msun, $d < 240$\,pc (116 systems)
    \item $0.6 \leq M_1 < 0.7$\,\msun, $d < 250$\,pc (151 systems)
    \item $0.7 \leq M_1 < 0.8$\,\msun, $d < 290$\,pc (229 systems)
\end{enumerate}
as our final samples (see Figure~\ref{fig:AMRF}). We note that because of the primary mass uncertainty, there is some overlap between the samples. A total of 314 systems have passed these selection criteria, and are listed in Table~\ref{tab:sample}.
The $0.5 \leq M_1 < 0.6$\,\msun\ sample is the only one that probes the full WD mass range, where the expected AMRF values of systems with $\gtrsim 0.2$\,\msun\ companions are higher than the maximal AMRF value of an MS+MS binary. The other two samples are biased towards more massive WDs, as $\lesssim 0.6$\,\msun\ WD companions in this primary mass range are expected to be classified as \textit{class-I} systems (see Figure~\ref{fig:AMRF}).

We also define a larger-yet-biased and volume-incomplete sample of systems with no red-color excess, for reference. This sample is selected from the 6,880 systems that have passed the selection criteria listed in Equations~(\ref{eq:PrI})--(\ref{eq:initial_m1max}), with the additional no red-color excess condition:
\begin{equation}
    \pr{\text{red}} \geq 0.56,
\end{equation}
resulting in 2,135 systems. Note that this condition applies both to \textit{class-II} and \textit{class-III} systems, unlike the one in Equation~(\ref{eq:NCEclassII}), to avoid the artificial excess of systems with $M_1 \approx 1$\,\msun\ (see above).

See Appendix~\ref{sec:properties} for the system properties of the different selected samples.

\section{Results} \label{sec:results}

\begin{figure}
    \centering
    \includegraphics[width=\columnwidth]{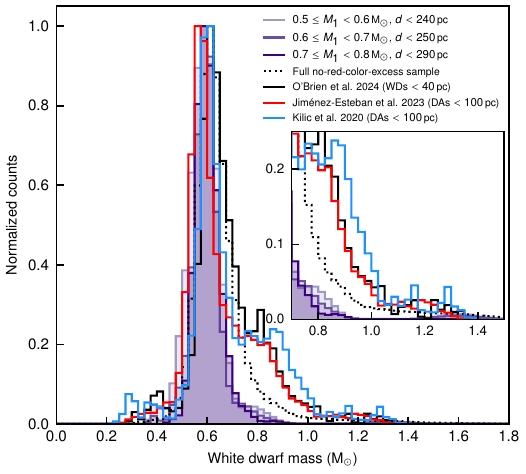}
    \caption{Secondary mass distribution of the selected volume-limited samples with MS primaries of $0.5 \leq M_1 < 0.6$\,\msun, $0.6 \leq M_1 < 0.7$\,\msun, and $0.7 \leq M_1 < 0.8$\,\msun\ (purple-shaded bars), and the full no-red-color-excess sample (dotted black).
    The observed distribution of the full 40\,pc WD sample of \citet[][solid black]{OBrien_2024}, and the magnitude-limited 100\,pc hydrogen-dominated (DA) WD samples of \citet[][solid red]{JimenezEsteban_2023} and \citet[][solid blue]{Kilic_2020} are plotted for reference. The inset panel zooms on the high-mass end of the distribution.}
    \label{fig:WDmass}
\end{figure}

Figure~\ref{fig:WDmass} shows the secondary mass distribution of our three volume-limited samples, as well as of our volume-incomplete no-red-color-excess sample. To take the mass uncertainty into account, we represent each star's contribution to the distribution with a Gaussian mass probability distribution having the measured mean and standard deviation of the secondary mass. We then calculate the secondary mass distribution of our sample by summing all of the Gaussians into mass bins between 0 and 1.8\,\msun, in 0.025\,\msun\ steps.
We compare our results to the observed mass distributions of single WDs in the field. We calculate these distributions in a similar way (summing Gaussians) using the published catalogs of \citet{OBrien_2024}, \citet{JimenezEsteban_2023}, and \citet{Kilic_2020}. Finally, we normalize each distribution by its maximal value, making the comparison between the height of the main $\approx 0.6$\,\msun\ peak and that of the high-mass bump insensitive to the inclusion of low-mass WDs.

The WD mass distribution of \citet{OBrien_2024} is based on the mass estimates of 1,051 out of 1,076 WDs in the volume-complete 40\,pc WD sample.\footnote{For 25 WDs in this sample the mass could not be determined reliably due to contamination from nearby stars or the presence of collision-induced absorption (CIA) or strong molecular carbon absorption bands \citep{OBrien_2024}. For an additional 0.57\,\msun\ WD (WD\,J181706.50+132824.99) we changed the mass uncertainty from the reported 0.00 to 0.001\,\msun, the same as that of another WD (WD\,J074538.43-335551.34) with a similar $\log g$ error.} The WD masses in this catalog were estimated by fitting their spectral energy distributions (SEDs) using theoretical WD models, after determining their spectral type spectroscopically. Theses masses were then corrected to account for inaccuracies in the low-mass models \citep{OBrien_2024}.

The samples of \citet{JimenezEsteban_2023} and \citet{Kilic_2020} are both based on the 100\,pc WD sample.
The WD mass distribution of \citet{JimenezEsteban_2023} is based on 5,728 hydrogen-dominated (DA) WDs hotter than 5,500\,K within 100\,pc. The WD mass distribution of \citet{Kilic_2020} is based on 1,351 DA WDs hotter than 6,000\,K within 100\,pc and the SDSS footprint, where their spectroscopically confirmed catalog is $83\%$ complete. The WD masses in both catalogs were also estimated by fitting their SEDs using theoretical WD models, after verifying spectroscopically that they have hydrogen-dominated atmospheres (using SDSS and follow-up spectra in \citealt{Kilic_2020}, and Gaia's low-resolution BP/RP spectra in \citealt{JimenezEsteban_2023}).

\begin{figure}
    \centering
    \includegraphics[width=\columnwidth]{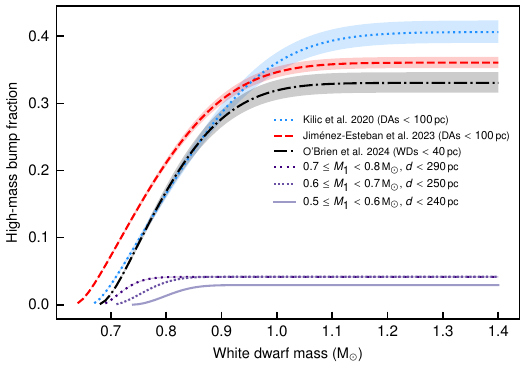}
    \caption{The cumulative fraction of massive WDs, relative to the full distribution. See Appendix~\ref{sec:gaussians} for the fitting procedure used to estimate the high-mass fraction).
    }
    \label{fig:HighMassFraction}
\end{figure}

Figure~\ref{fig:HighMassFraction} shows the cumulative fraction of massive WDs relative to the full distribution of each sample (see Appendix~\ref{sec:gaussians} for the fitting procedure used to estimate the high-mass fraction). We find that while the high-mass bump constitutes a significant fraction of the 100\,pc field-WD sample---$40.6\pm1.7\%$ in the sample of \citet{Kilic_2020}, $36.1\pm0.8\%$ in that of \citet{JimenezEsteban_2023}, and $33.1\pm1.6\%$ in that of \citet{OBrien_2024}---it contributes as much as $2.92\pm0.02\%$ ($4.16\pm0.04\%$) in our sample of WDs in astrometric binaries with $0.5 \leq M_1 < 0.6$\,\msun\ ($0.6 \leq M_1 < 0.7$\,\msun)\footnote{No high-mass bump was detected for the $0.7 \leq M_1 < 0.8$\,\msun, see Appendix~\ref{sec:gaussians}.}. In other words, the fraction of WDs more massive than $\sim 0.7$\,\msun\ in binaries with K/M-dwarf primaries and orbital separations $\sim 1$\,AU, is smaller by a factor $\sim 8-14$ compared to their fraction among single WDs in the field.

\section{Discussion}\label{sec:discussion}

Figures~\ref{fig:WDmass} and \ref{fig:HighMassFraction} show a clear deficit of massive WDs in all of our MS+WD samples compared to the field WD samples. This is despite the bias \emph{against} massive WDs in the magnitude-limited 100\,pc field samples on the one hand (see Section~\ref{sec:intro}), and despite Gaia's higher observational sensitivity to astrometric binaries with more massive companions on the other hand. Such systems have higher AMRF values, which facilitate their detection and identification (see Figure~\ref{fig:AMRF}).
This deficit could arise from a selection effect, from an intrinsic physical difference between these two populations, or from a combination of the two.

\subsection{Post-AGB-interaction binaries}

The orbital separation of $\sim 1$\,AU indicates that binary interaction of some kind was unavoidable when the WD was still in its red-giant phase. At the same time, it rules out a common-envelope phase, that should have brought the two stars much closer to each other. This suggests stable mass transfer as the interacting mechanism. The large eccentricities of some of the systems in our sample (see Figure~\ref{fig:EccentricityVsPeriod} in Appendix~\ref{sec:properties}) require an eccentricity pumping mechanism to counteract the circularization induced by the stable mass transfer (e.g.~\citealt{BonacicMarinovic_2008, VanWinckel_1995, Waelkens_1996, Soker_2000, Izzard_2010, Dermine_2013, Vos_2015}; and see discussion in \citealt{Shahaf_2024}).

\begin{figure}
    \centering
    \includegraphics[width=\columnwidth]{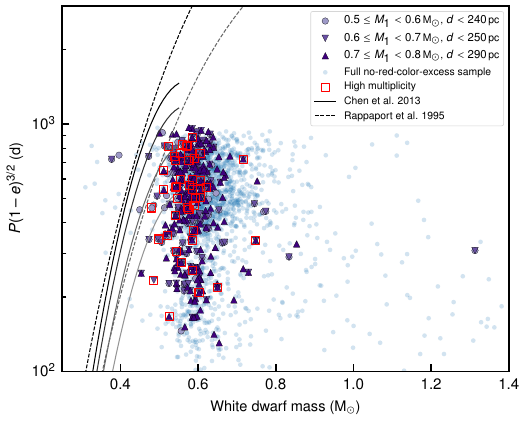}
    \caption{Mass-period relation of the selected samples (purple markers indicating different primary mass ranges as in Figure~\ref{fig:AMRF}, see legend), and the full no-red-color-excess sample (blue dots). Systems of higher multiplicity (see Section~\ref{sec:tertiary}) are marked by a red square. The solid lines show the theoretical relation from \citet{Chen_2013} for different metallicities (from top to bottom, $Z=0.02,~0.01,~0.004,~0.001$), while the dashed lines show the relation from \citet{Rappaport_1995} for $Z=0.02$ and $0.004$. The $y$-axis shows the orbital period, $P$, times $(1-e)^{3/2}$, where $e$ is the eccentricity \citep[see][for a derivation of this relation]{Joss_1987}.}
    \label{fig:MassPeriod}
\end{figure}

Systems that have undergone stable mass transfer are expected to have some kind of a correlation between the mass of the WD and the orbital period \citep[e.g.][]{Joss_1987, Rappaport_1995, Chen_2013}. These relations have been calculated and tested for systems with low-mass WDs ($\lesssim 0.55$\,\msun), i.e.~when the WD progenitor was on the red-giant branch (RGB). However, the relatively large masses of the majority of the WDs in our sample indicate that this stable mass transfer started when the WD progenitor was already on the asymptotic giant branch (AGB). Indeed, the mass-period relation of our observed sample is substantially different than that predicted for post-RGB interaction, with massive WDs in much shorter orbits than expected (Figure~\ref{fig:MassPeriod}). Yet it is still likely that some kind of a mass-period relation does exist for post-AGB systems as well. Such a relation might impose an implicit prior on the WD mass distribution of our sample, given the pre-selected orbital separation range (determined by Gaia's astrometric sensitivity).

The interaction on the AGB and the period-eccentricity relation of the systems in our sample resemble those of barium stars. Although a few known barium stars are found in the full MS+WD sample of \citet{Shahaf_2024}, none of them is part of the subsamples used here. This is not surprising, as known barium stars are usually more massive than the MS primaries included in our samples.
\citet{Merle_2016} have analyzed the \textit{s}-process element abundance of a small sample of 11 WDs with measured masses, orbiting cool ($<7,000$\,K) stars. They showed that WDs less massive than 0.5\,\msun\ do not have barium-star companions. This is expected since the progenitors of $<0.5$\,\msun\ WDs could not have reached the thermally pulsing-AGB (TP-AGB) phase \citep{Hurley_2000}.
\citet{Kong_2018} and \citet{BharatKumar_2019} arrived at similar conclusions based on a sample of 18 and 21 Sirius-like systems, respectively. For WDs with masses $>0.51$\,\msun\ the results were inconclusive: some of the companions to massive WDs did not show evidence of \textit{s}-process element enrichment, demonstrating that the mass of the WD progenitor is not the only parameter responsible for the creation of barium stars (with the other main suspects being the orbital separation and metallicity).

\begin{figure}
    \centering
    \includegraphics[width=\columnwidth]{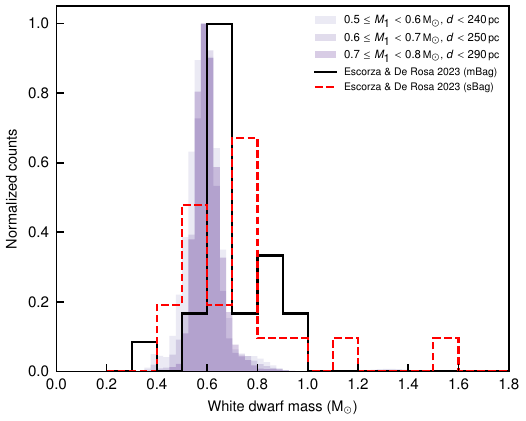}
    \caption{The WD mass distributions of our samples (purple-shaded bars) compared to that of WD orbiting mild barium giants (black solid line) and strong barium giants (red dashed line) from \citet{Escorza_2023}.}
    \label{fig:WDmassBaStars}
\end{figure}

Recently, \citet{Escorza_2023} derived the mass distribution of WD companions to barium stars. They found a possible relation between the \textit{s}-process element abundance in the barium red-giant primary and the WD mass, where strongly polluted barium giants (`sBag') tend to have more massive WD companions than mildly polluted barium giants (`mBag'). However, this relation was not sufficiently statistically significant (with a p-value of 0.45 in a Kolmogorov-Smirnov test).
Figure~\ref{fig:WDmassBaStars} compares the mass distributions of the WD companions of strong and mild barium giants from \citet{Escorza_2023}, and those of our volume-limited samples. Note that the samples of \citet{Escorza_2023} are not volume complete, and their mass distributions presented in Figure~\ref{fig:WDmassBaStars} are regular histograms (normalized by the peak of the mild barium giant sample), and not the sum of Gaussians (see Section~\ref{sec:results}).

Figure~\ref{fig:WDmassBaStars} shows that the WD mass distributions of our samples peak at a slightly lower value than that of WDs orbiting mild and strong barium giants. A Kolmogorov-Smirnov test between our samples and the barium samples yielded p-values $\lesssim 5 \times 10^{-9}$ for the mild barium giants, and $\lesssim 10^{-4}$ for the strong barium giants, indicating a statistically significant difference between the samples. Assuming that the systems in our sample are post-AGB binaries, this might hint at a possible `continuum' of \textit{s}-process element pollution, depending on the WD progenitor mass. In an ongoing follow-up program, we obtain mid-resolution spectra of systems in the AMRF sample \citep{Shahaf_2024}. If this relation is correct, we expect to see a correlation between the \textit{s}-process element abundance in the MS spectra, and the mass of their WD companions \citep[as suggested by][]{Boffin_1988}. Theoretical calculations by \citet{Karakas_2016} indicate that stars with initial masses of $2.5-4$\,\msun\ yield more \textit{s}-process elements than lower- and higher-mass stars. In this case, the MS companions to the less-massive WDs in our sample might lack any traces of \textit{s}-process elements within the observational sensitivity. Since current post-AGB binary samples focus on barium stars, they can miss the lower- or higher-mass end of the underlying population. Gaia's astrometric binary sample thus offers a more complete view of the population of post-AGB binaries.

If our samples represent indeed a population of post-stable-mass-transfer systems, then our primary mass selection implies an indirect prior on the initial mass ratio of the systems, since stable mass transfer is unlikely in case of extreme mass ratios of the progenitor systems \citep{Hjellming_1987, Hurley_2002}. Thus, the deficit of massive WDs in our selected samples can be the result of the selection of the primary mass ($0.5-0.8$\,\msun) and the orbital separation range probed by Gaia (which implies post-stable-mass-transfer binaries).

Yet when looking at our three different volume-limited samples, we see no significant difference between their resulting WD mass distributions (Figure~\ref{fig:WDmass}), despite the different primary mass distributions. In other words, their present-day mass-ratio distribution is different (Figure~\ref{fig:SystemProperties} in Appendix~\ref{sec:properties}).
Moreover, even when looking at the full no-red-color excess sample, there is still a clear deficit of massive WDs compared to the field samples (see dotted black line in Figure~\ref{fig:WDmass}). This sample contains MS primaries as massive as $\sim 1.2$\,\msun\ (Figure~\ref{fig:SystemProperties}), thus it does not exclude a priori systems with more massive WD companions. In addition, since this sample is not volume-limited, it is biased in favor of more massive MS primaries (due to their higher luminosities), and as a consequence---of more massive WD companions. Furthermore, by excluding \textit{class-I} systems (Equation~\ref{eq:PrI}), we exclude MS+WD systems with $\sim 0.6$\,\msun\ WDs and $\gtrsim 0.8$\,\msun\ MS primaries (see Figure~\ref{fig:AMRF}). Since we normalize the WD mass distribution by the $\sim 0.6$\,\msun\ peak, the bias against $\sim 0.6$\,\msun\ WDs artificially increases the share of massive WDs in the normalized distribution. Yet despite these initial conditions biased in favor of massive WDs, there is still a lack of massive WDs in this sample compared to the field samples of WDs.
Hence, the implicit prior on the initial mass ratio of the system does not seem to be the only reason for the massive-WD deficit.

\subsection{Lack of merger products}\label{sec:mergers}

There is growing evidence that a very large fraction of massive WDs are in fact merger products of binary evolution rather than descendants of single stars. There is a significantly smaller fraction of photospheric metal pollution from accretion of planetary debris in massive WDs compared to canonical-mass WDs \citep{Koester_2014}. This could be explained by the complicated past of a merger-product WD, that makes it unlikely for planetary systems to survive the process.
Other indications are strong magnetic fields, rapid rotation rates, and high transverse velocities of many massive WDs \citep[e.g.][]{Kilic_2023}. The lack of massive WDs in the mass distribution of helium-dominated (DB) WDs \citep{Bergeron_2001, GentileFusillo_2019, Tremblay_2019} also agrees with the merger origin of massive WDs, as only rare mergers of DB+DB binaries will produce a massive DB WD.
\citet{Maoz_2018} have estimated the double-WD merger rate based on a large spectroscopic sample of WDs. They found that the merger rate is $4.5-7$ times higher than the specific Type Ia supernova rate in the Milky Way, implying that a large fraction of present-day WDs should be merger products. Based on transverse velocities from Gaia, \citet{Cheng_2020} have estimated that about $20\%$ of massive ($\gtrsim0.8$\,\msun) WDs have dynamical ages that are inconsistent with their cooling ages, indicating a merger origin. \citet{Fleury_2022} have estimated that $\approx 40-50\%$ of WDs with masses of $1.15-1.25$\,\msun\ come from mergers, by comparing their cooling age distribution with the star formation history.
Theoretical binary population synthesis studies have also estimated a significant contribution of mergers to the currently single WD population, of about $10-30\%$ \citep{Toonen_2017} or $9-16\%$ \citep{Torres_2022} of all single WDs, and $30-50\%$ of the massive WDs \citep{Temmink_2020}.
Interestingly, there might be a hint of an additional small peak around $\approx 1.2$\,\msun\ in the mass distribution of field WDs, as can be seen in Figure~\ref{fig:WDmass}. If this excess of $\approx 1.2$\,\msun\ WDs is genuine indeed, it might be the result of equal-mass double-WD mergers (of two $\approx 0.6$\,\msun\ WDs). This would be consistent with the ``twin'' peak discovered by \citet{ElBadry_2019} in wide MS binaries from Gaia.

Given the relatively small orbital separations ($\sim 1$\,AU) of the MS+WD binaries in our sample, it is unlikely that a large number of the WD companions in these systems (if any) are merger products of an inner binary in a previously hierarchical triple system, since the hypothetical inner binary could not have had enough room to evolve and merge.
In the most extreme case, where the difference between our samples and field samples stems only from the lack of merger products, it implies that almost all WDs more massive than $\sim 0.7$\,\msun\ come from mergers. However, since we have no accurate knowledge of the initial mass-ratio distribution of our sample and of the stable mass-transfer mechanism these systems have undergone, we cannot confidently quantify the contribution of mergers compared to the field samples.

\subsection{Searching for tertiary companions}\label{sec:tertiary}
Recently, \citet{Shariat_2023} suggested that at least $30\%$ of solar-type stars in the local neighborhood were born in hierarchical triples, based on simulations and on the Gaia 200\,pc sample. From their simulations \citet{Shariat_2023} find that the inner binary in the majority of these triples merge and that only a few per cents remain in an MS+WD configuration within the Galaxy lifetime. Since the orbital separations of the inner binaries in those systems are typically on the order of $\sim 1$\,AU as well, we have searched the Gaia DR3 catalog for tertiary companions within a projected separation of 1\,pc around the MS+WD binaries in our selected subsamples. See Appendix~\ref{sec:tertiarysearch} for the details of this search. We find tertiary companions to 44 out of our 314 systems ($\approx 14\%$), as listed in Table~\ref{tab:tertiary}, in consistency with \citet{Shariat_2023}. This is also consistent with the lack of merger products assumption (see Section~\ref{sec:mergers}), since the systems with merged inner binaries have much larger orbital separations.

The eccentricities of the inner binaries in these systems are not different than those of the non-triple systems in our sample (Figure~\ref{fig:EccentricityVsPeriod}), and are somewhat lower than those of the simulated systems of \citet[][C. Shariat, private communication]{Shariat_2023}.
The tertiary companions are almost always less massive the the primary MS in the inner binary (see Figure~\ref{fig:TertiariesColor}). This might be a selection effect, since in systems with more massive tertiaries, the tertiary is more likely to have evolved into a compact object, making it harder to detect. A few of these systems might actually be quadruple systems, composed of an inner MS+WD binary and an outer MS+WD binary companion, based on their location on the HR diagram between the MS and the WD sequence (see Figure~\ref{fig:Tertiaries}).
Triple systems in which the tertiary is an evolved star---especially a single WD---are of particular interest, since the tertiary can be used to estimate the total age of the system, and thus constrain the binary evolution and initial-to-final mass relation of the WD in the inner binary. In our subsample there is three such system with WD tertiaries. However, in two of them the tertiary seems to be an extremely low-mass WD, that could not have evolved from a single star in the Universe lifetime. We defer the analysis of these systems and an expanded search for tertiary companions in the full MS+WD catalog of \citet{Shahaf_2024} to an upcoming publication.

\section{Summary and conclusions}
In this study, we find a stark deficit of massive WD companions in volume-complete samples of binaries with K/M-dwarf primaries and orbital separations of $\sim 1$\,AU, compared to their occurrence among single WDs in the field. These systems are likely to have undergone a phase of stable mass transfer when the WD progenitor was on the AGB. This might indicate an implicit mass-ratio and/or WD-mass prior selection explaining (part of) this deficit. In addition, given the relatively small orbital separations, these systems are unlikely to contain WDs that are merger products.

These subsamples, as well as the full sample from which they were drawn \citep{Shahaf_2023, Shahaf_2024}, provide a unique opportunity to study the population of post-AGB-interaction binaries (barium stars included), which is currently poorly constrained by observations. The discovery of stable mass-transfer WD+M binaries with orbital periods of a few hundred days contradicts previous assumptions on the possible existence of such systems \citep{Lagos_2022}.
In addition, it adds to the accumulating evidence that a large fraction of the massive WDs in the field are merger products. This has implications to many astrophysical and cosmological questions. The merger rate and mass distribution of WDs directly affect the expected rate of Type-Ia supernovae \citep[assuming double-degenerate scenarios, e.g.][]{Maoz_2018}, and the expected gravitational-wave foreground and detectable sources for the Laser Interferometer Space Antenna \citep[LISA;][]{Korol_2022}. It also affects the expected occurrence rate of massive-WD companions in short-period binaries with compact companions \citep[e.g.][]{Mazeh_2022}.

\section{Acknowledgments}

We thank Boris G\"{a}nsicke and Dan Maoz for pointing out that the narrow distribution and lack of a high-mass tail in the mass distribution of DB WDs hints at a merger origin, Cheyanne Shariat for suggesting that some of the systems might have tertiary companions, and the anonymous referee for valuable comments.

The research of SS is supported by a Benoziyo prize postdoctoral fellowship.
SBA and TM acknowledge support from the Israel Ministry of Science grant IMOS 714193-02.
SBA's research is supported by the Peter and Patricia Gruber Award; the Azrieli Foundation; the Andr\'e Deloro Institute for Advanced Research in Space and Optics; the Willner Family Leadership Institute for the Weizmann Institute of Science; and the Israel Science Foundation grant ISF 714022-02.
SBA is the incumbent of the Aryeh and Ido Dissentshik Career Development Chair.

This work has made use of data from the European Space Agency (ESA) mission Gaia (\url{https://www.cosmos.esa.int/gaia}), processed by the Gaia Data Processing and Analysis Consortium (DPAC; \url{https://www.cosmos.esa.int/web/gaia/dpac/consortium}). Funding for the DPAC has been provided by national institutions, in particular the institutions participating in the Gaia Multilateral Agreement.

%

\vspace{5mm}
\facilities{Gaia}


\software{astropy \citep{Astropy_2013, Astropy_2018}, matplotlib \citep{Hunter_2007}, numpy \citep{Numpy_2006, Numpy_2011}, scipy \citep{Virtanen_2020}, topcat \citep{Taylor_2005}
          }



\appendix

\section{Sample properties}\label{sec:properties}

\begin{figure*}
    \centering
    \includegraphics[width=\textwidth]{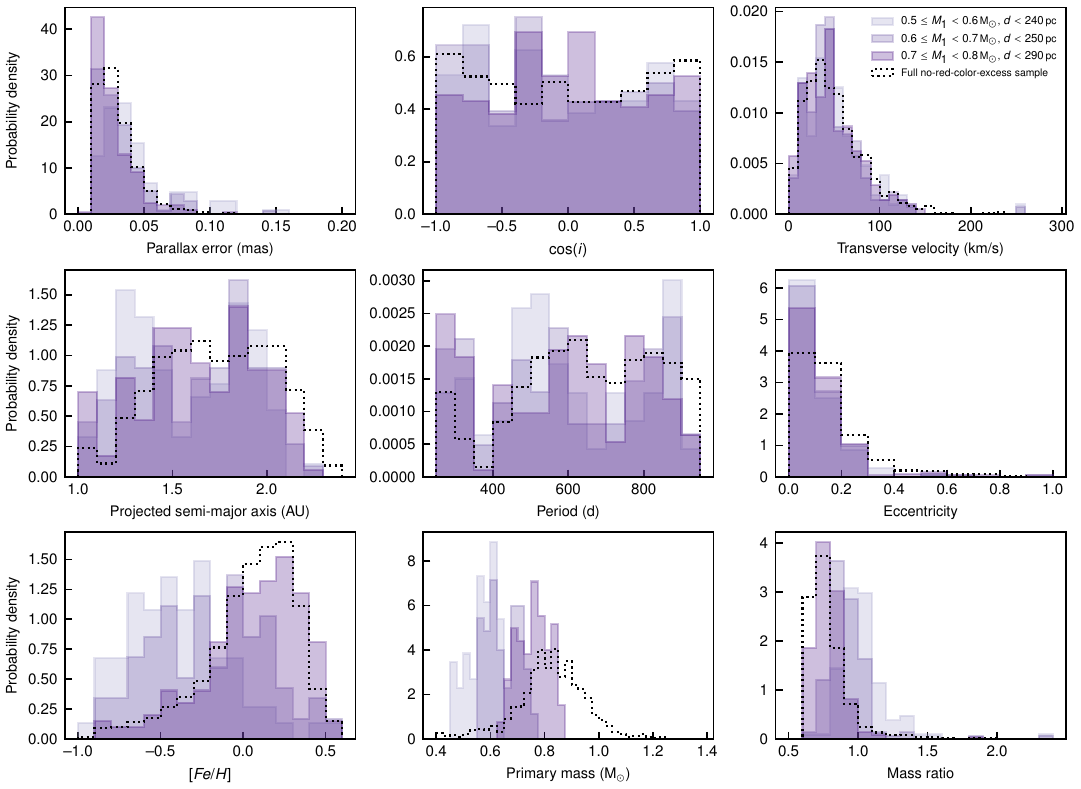}
    \caption{Histograms of the parallax error, cosine of the inclination angle, transverse velocity, projected semi-major axis, period, eccentricity, metallicity, primary mass, and mass ratio (\textit{from right to left, top to bottom}) of the selected volume-limited samples and the reference full no-red-color-excess sample. The dip in the period around 2yr is a result of Gaia's mission duration.}
    \label{fig:SystemProperties}
\end{figure*}

\begin{figure}
    \centering
    \includegraphics[width=\columnwidth]{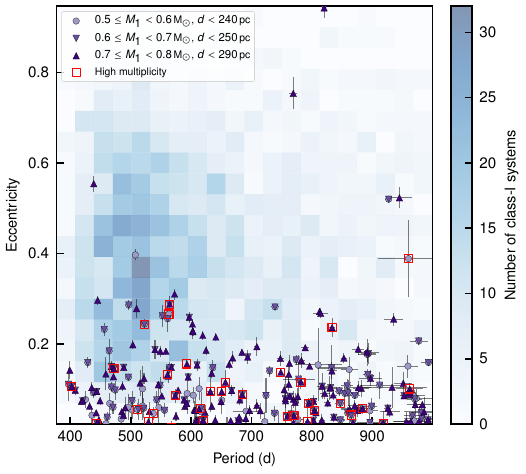}
    \caption{Period-eccentricity relation of the selected samples (purple markers with black error bars). Systems of higher multiplicity (see Section~\ref{sec:tertiary}) are marked by a red square. The two-dimensional histogram of a \textit{class-I} ($\Pr{I} > 0.9$) subsample with the same constraints on the primary mass, orbital separation, and distance is plotted in blue shades for reference.}
    \label{fig:EccentricityVsPeriod}
\end{figure}

Figure~\ref{fig:SystemProperties} shows the parallax error, inclination, transverse velocity, projected semi-major axis, period, eccentricity, metallicity, primary mass, and mass ratio distributions of our volume-limited samples, along with the full no-red-excess sample, for reference. No significant difference is seen between these samples.
The estimated metallicities of systems with $\lesssim 0.6$\,\msun\ primaries are generally unreliable \citep[see][]{Shahaf_2024}, explaining the different metallicity distributions of the $0.5 \leq M_1 < 0.7$\,\msun\ subsamples.
The eccentricity as a function of the orbital period is shown in Figure~\ref{fig:EccentricityVsPeriod} (see discussion in Section~\ref{sec:discussion}).
Table~\ref{tab:sample} includes the full selected sample (all the 314 systems that have passed the selection criteria listed in Section~\ref{sec:sampleselection}).

\begin{deluxetable*}{cccccccc}\label{tab:sample}
\tablecaption{The selected sample. $\varpi$ is the parallax, $M_1$ is the mass of the MS component, $M_2$ is the mass of the WD, $P$ is the orbital period, and $e$ is the eccentricity. The full table is available in the supplemental information accompanying this publication.}
\tablehead{\colhead{Gaia DR3 ID} & \colhead{RA} & \colhead{Dec} & \colhead{$\varpi$} & \colhead{$M_1$} & \colhead{$M_2$}& \colhead{$P$} & \colhead{$e$} \\ 
\colhead{} & \colhead{(deg)} & \colhead{(deg)} & \colhead{(mas)} & \colhead{(\msun)} & \colhead{(\msun)} & \colhead{(d)} & \colhead{} } 
\startdata
3598782754069068032 & 176.2898023 & -5.7129768 & $4.801\pm0.048$ & $0.452_{-0.051}^{+0.049}$ & $0.635\pm0.043$ & $862\pm14$ & $0.144\pm0.027$\\
5299929002634004480 & 136.5975904 & -60.7005195 & $6.394\pm0.042$ & $0.453_{-0.051}^{+0.049}$ & $0.594\pm0.036$ & $507.7\pm1.4$ & $0.397\pm0.012$\\
868863951177644160 & 115.5600885 & 25.9453065 & $5.72\pm0.11$ & $0.456_{-0.050}^{+0.051}$ & $0.555\pm0.043$ & $465.2\pm2.4$ & $0.069\pm0.024$\\
2856698098507352064 & 3.4980152 & 26.9011924 & $4.217\pm0.073$ & $0.458_{-0.050}^{+0.050}$ & $0.519\pm0.036$ & $593.5\pm5.5$ & $0.091\pm0.045$\\
4263071704959743360 & 288.9649287 & -0.8243133 & $4.36\pm0.10$ & $0.460_{-0.051}^{+0.050}$ & $0.627\pm0.053$ & $546.7\pm4.9$ & $0.060\pm0.053$\\
2566461354152574976 & 21.0477577 & 7.9769125 & $5.33\pm0.11$ & $0.461_{-0.050}^{+0.051}$ & $0.553\pm0.044$ & $616.7\pm3.5$ & $0.051\pm0.043$\\
224549450109569536 & 54.6271642 & 39.2316386 & $7.576\pm0.051$ & $0.465_{-0.050}^{+0.050}$ & $0.514\pm0.030$ & $484.24\pm0.93$ & $0.035\pm0.011$\\
6685604337007194368 & 300.3137539 & -44.6334559 & $4.936\pm0.037$ & $0.472_{-0.050}^{+0.050}$ & $0.486\pm0.028$ & $519.5\pm2.3$ & $0.064\pm0.032$\\
6218848456877200640 & 220.6254098 & -29.3426568 & $4.62\pm0.16$ & $0.473_{-0.050}^{+0.051}$ & $0.521\pm0.048$ & $810\pm39$ & $0.154\pm0.080$\\
159844874438662400 & 70.3915029 & 32.2002141 & $10.467\pm0.044$ & $0.475_{-0.050}^{+0.050}$ & $0.630\pm0.042$ & $973\pm25$ & $0.080\pm0.020$\\
\enddata
\end{deluxetable*}

\section{Estimating the high-mass fraction}\label{sec:gaussians}

\begin{figure}
    \centering
    \includegraphics[width=\columnwidth]{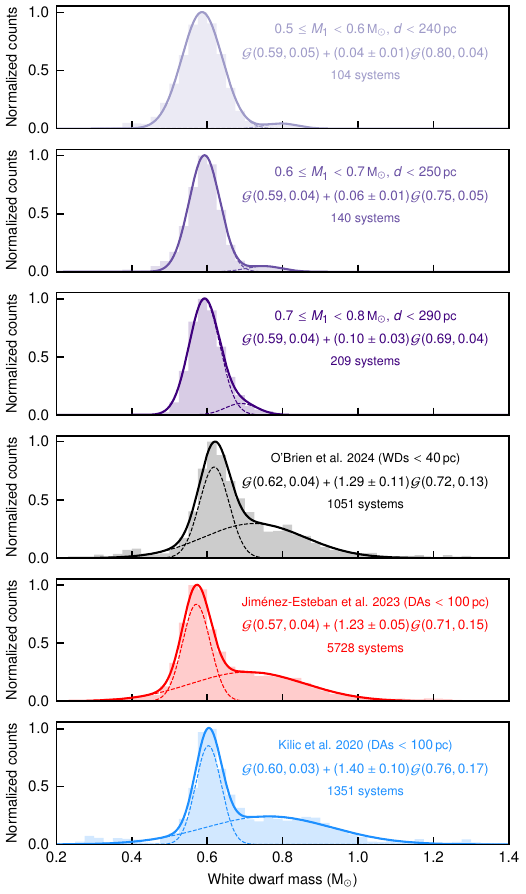}
    \caption{Double-Gaussian fit to the mass distribution of our final samples (\textit{top three panels}), the 40\,pc dield sample of \citet{OBrien_2024} (fourth panel), and the 100\,pc field samples of \citet{JimenezEsteban_2023} (\textit{fifth panel}), and \citet{Kilic_2020} (\textit{bottom panel}). The bar plot shows the underlying observed distribution. The solid line shows the fitted function, while the dashed lines show the individual Gaussian components of the fit.}
    \label{fig:GaussianFit}
\end{figure}

In order to compare the different distributions, we fit each sample with a combination of two Gaussians, $\mathcal{G}(\mu_1, \sigma_1) + n\mathcal{G}(\mu_2, \sigma_2)$ normalized by the maximal peak, where
\begin{equation}
    \mathcal{G} \left( \mu, \sigma \right) = \frac{1}{\sigma\sqrt{2\pi}} e^{-\frac{\left(x-\mu\right)^2}{2\sigma^2}}
\end{equation}
is a Gaussian with mean $\mu$ and standard deviation $\sigma$, and $n$ is a constant. We fit the distribution using \textsc{scipy}'s \textsc{curve\_fit} Trust Region Reflective algorithm, setting the initial positions of the Gaussians' means to 0.6 and 0.8\,\msun, and the lower bound of the second Gaussian mean to 0.69\,\msun.
Figure~\ref{fig:GaussianFit} shows the fit results for each of the samples. Note that no actual secondary peak is detected for the $0.7 \leq M_1 < 0.8$\,\msun\ sample, and the secondary Gaussian's mean is assigned the lower limiting value allowed for the fit. We thus exclude the Gaussian fit of this sample from the discussion.

We define the area enclosed between the two Gaussians, integrated from their intersection point and normalized by the area enclosed by the full double-Gaussian distribution, as the fraction of massive WDs in each sample. Figure~\ref{fig:HighMassFraction} shows the cumulative fraction of massive WDs. The shaded area in Figure~\ref{fig:HighMassFraction} marks the uncertainty based on the $1\sigma$ error of the $n$ coefficient (see Figure~\ref{fig:GaussianFit}).

\section{Tertiary search}\label{sec:tertiarysearch}

\begin{figure}
    \centering
    \includegraphics[width=\columnwidth]{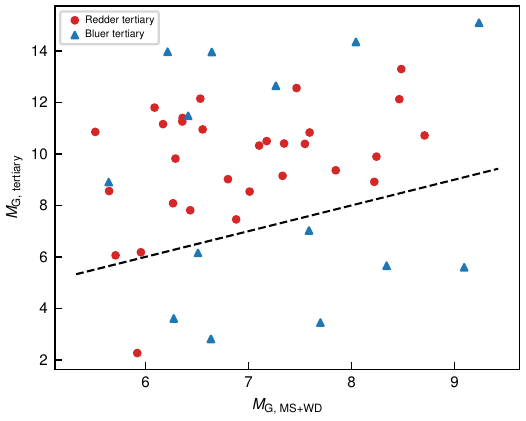}
    \caption{Dereddened absolute Gaia $G$-band magnitude of the tertiary companions vs that of the corresponding inner binary. Systems where the tertiary has a redder (bluer) $G_\text{BP}-G_\text{RP}$ color are marked by red circles (blue triangles). The dashed black line corresponds to equal absolute magnitudes.}
    \label{fig:TertiariesColor}
\end{figure}

\begin{figure}
    \centering
    \includegraphics[width=\columnwidth]{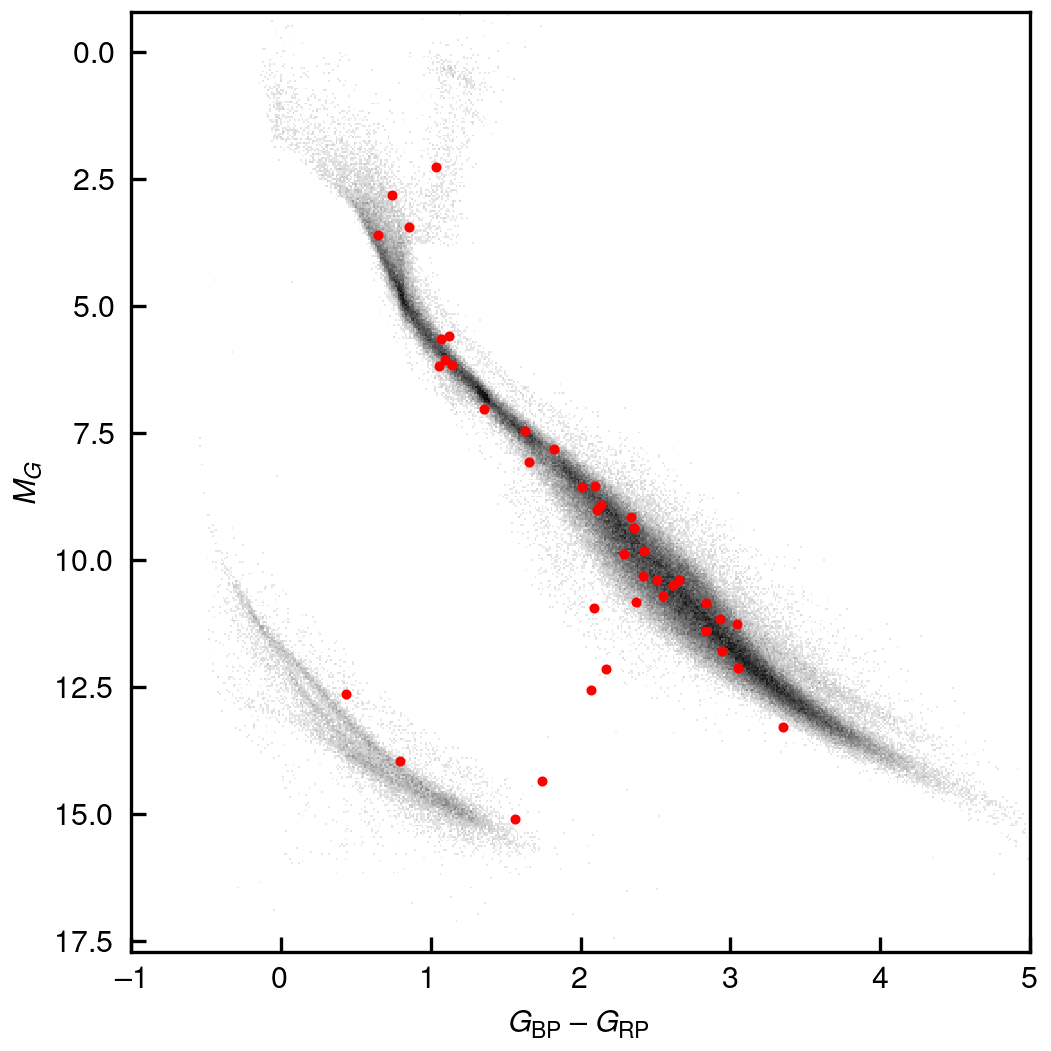}
    \caption{Gaia dereddened color-magnitude diagram for the tertiary companions in the higher-multiplicity systems in our sample (marked by a red circle). The number density of the Gaia 100\,pc sample is shown in grayscale for reference.}
    \label{fig:Tertiaries}
\end{figure}

We searched the single-star catalog of Gaia DR3 (\texttt{gaiadr3.gaia\_source}) for tertiary companions within a projected separation of 1\,pc. We followed \citet{ElBadry_2021}, selecting sources with valid parallaxes (larger than 1\,mas, with single-to-noise ratio larger than 5 and error smaller than 2\,mas) and non-null $G$-band photometry. We further selected only sources with consistent parallaxes and proper motions (equations 1--7 in \citealt{ElBadry_2021}). We find 44 systems with tertiary companions. The results are listed in Table~\ref{tab:tertiary}. Figure~\ref{fig:TertiariesColor} compares the absolute magnitudes and colors of the tertiaries and their corresponding inner binaries. In most cases the tertiaries are fainter and redder than the inner binaries, indicating a lower mass compared with the primary MS stars. Figure~\ref{fig:Tertiaries} shows the location of these tertiaries on the Gaia color-magnitude diagram.

\begin{deluxetable*}{ccccc}\label{tab:tertiary}
\tablecaption{The high-multiplicity systems in our sample. The first ID is of the MS+WD inner binary, while the second ID and the coordinates are of the outer companion. $a_\text{out}$ is the projected orbital separation of the outer companion.}
\tablehead{\colhead{Gaia DR3 ID$_1$} & \colhead{Gaia DR3 ID$_2$} & \colhead{RA} & \colhead{Dec} & \colhead{$a_\text{out}$} \\ 
\colhead{} & \colhead{} & \colhead{(deg)} & \colhead{(deg)} & \colhead{(AU)}} 
\startdata
41954481793705984 & 41947163169437312 & 51.3936258 & 14.1095859 & $\left(6.449 \pm 0.067\right) \times 10^{4}$\\
613025599896017152 & 613025595603558656 & 146.2528781 & 11.6631120 & $\left(2.7 \pm 1.4\right) \times 10^{2}$\\
811943986554770944 & 811943986554770304 & 140.0323833 & 38.4901015 & $7793 \pm 67$\\
1464711807899087872 & 1464710948905628544 & 195.7554838 & 30.7102862 & $4063 \pm 25$\\
1483641103161355520 & 1483641103161355648 & 211.5761748 & 37.6017495 & $1812 \pm 41$\\
1492136445393915008 & 1492136445393253760 & 214.9042418 & 41.5628127 & $\left(2.25 \pm 0.45\right) \times 10^{3}$\\
1538567721922127872 & 1538567721921601024 & 185.1857278 & 43.9048691 & $\left(5.3 \pm 1.6\right) \times 10^{2}$\\
1674635034638596608 & 1674635034638596480 & 206.7410714 & 70.1170684 & $4016 \pm 63$\\
1860726205116568960 & 1857719113838086528 & 309.2080331 & 29.6135867 & $\left(1.0958 \pm 0.0036\right) \times 10^{5}$\\
1923185170017264768 & 1923185170014670720 & 352.3316075 & 39.7634623 & $\left(1.43 \pm 0.40\right) \times 10^{3}$\\
1966726242802331008 & 1966726238496845184 & 324.2892688 & 41.8937876 & $\left(3.94 \pm 0.16\right) \times 10^{3}$\\
2009304139089565824 & 2009304143391973120 & 349.6644490 & 57.2390633 & $473 \pm 45$\\
2038294210577929216 & 2038293798261077120 & 288.9527488 & 29.0814771 & $12559 \pm 39$\\
2151393206408068224 & 2151394683876818560 & 275.0641460 & 57.3197284 & $1435 \pm 54$\\
2200473069016427392 & 2200473069009862912 & 337.4862905 & 59.5666542 & $788 \pm 73$\\
2343910098928012928 & 2343910098928013056 & 10.7852972 & -25.7370659 & $902 \pm 30$\\
2605394231962904064 & 2605395090956499968 & 346.2535778 & -11.3645117 & $\left(1.426 \pm 0.011\right) \times 10^{4}$\\
2673760597963293824 & 2673760593670425344 & 327.0004357 & -2.5428920 & $\left(7.7 \pm 1.2\right) \times 10^{2}$\\
2682585347007232640 & 2682585342712253440 & 333.0021960 & 2.0869376 & $4495 \pm 84$\\
3197547046716036480 & 3197547145498930432 & 61.8850833 & -5.0470784 & $\left(4.72 \pm 0.32\right) \times 10^{3}$\\
3363639445008582272 & 3363639440711596416 & 109.7579607 & 19.7387912 & $\left(4.33 \pm 0.20\right) \times 10^{3}$\\
3466095099578873600 & 3466095099578872960 & 181.6494833 & -33.8798655 & $1888 \pm 42$\\
4037114609474235264 & 4037114570697892224 & 270.8505810 & -37.1188976 & $5445 \pm 49$\\
4069921356289901312 & 4069921356279221120 & 270.4503637 & -22.1154815 & $\left(2.39 \pm 0.16\right) \times 10^{3}$\\
4401090925658852352 & 4401090547701727616 & 234.3723657 & -4.2793397 & $14516 \pm 56$\\
4465538043807424384 & 4465538043805350656 & 247.8194402 & 15.6409156 & $\left(9.6 \pm 1.6\right) \times 10^{2}$\\
4640209037975567744 & 4640209037975567488 & 36.3436481 & -75.3109028 & $2225 \pm 46$\\
4810205969559395584 & 4810205973855745920 & 75.1466966 & -46.6567928 & $1651 \pm 21$\\
4983401560858949248 & 4983401560858949120 & 19.3615182 & -43.7020454 & $424 \pm 13$\\
5015634534500904960 & 5015634530204196608 & 23.0808196 & -32.9924135 & $\left(3.33 \pm 0.10\right) \times 10^{3}$\\
5092889726161772416 & 5092890104118894080 & 63.0466602 & -19.1807261 & $4811 \pm 42$\\
5491104349224057984 & 5491104353520148608 & 105.5992774 & -55.1508980 & $2743 \pm 25$\\
5608254942340292352 & 5608254946640301696 & 103.3887846 & -30.4734182 & $2518 \pm 32$\\
5790161689395101696 & 5790161689388304896 & 212.9604669 & -75.4187154 & $7275 \pm 23$\\
5874328960960829312 & 5874328956628761216 & 222.1666976 & -62.6258238 & $1143 \pm 81$\\
5882977513221448704 & 5882977513221446272 & 232.5871052 & -57.8725815 & $2492 \pm 21$\\
5938730445019006592 & 5938762129005006720 & 256.7395782 & -48.0432136 & $\left(1.903 \pm 0.011\right) \times 10^{5}$\\
5979878155858301440 & 5979877679168231296 & 257.9089066 & -32.8234753 & $\left(4.015 \pm 0.069\right) \times 10^{4}$\\
6385252016957373824 & 6385252016957373952 & 338.6680990 & -68.9519637 & $8619 \pm 26$\\
6396056127449928832 & 6396056127449928704 & 328.7035022 & -69.2339582 & $4281 \pm 43$\\
6542671158289237248 & 6542671153994956800 & 345.1808744 & -42.4306665 & $773 \pm 59$\\
6593522402843124992 & 6593522368483386240 & 338.1420104 & -39.9830140 & $6493 \pm 91$\\
6784523000111321216 & 6784523000111320960 & 319.7664536 & -30.8980630 & $1522 \pm 75$\\
6789344324239269760 & 6789343907626004608 & 321.2985621 & -28.9285655 & $\left(7.25 \pm 0.16\right) \times 10^{4}$\\
\enddata
\end{deluxetable*}


\clearpage

\bibliography{MSWD_AMRF}{}
\bibliographystyle{aasjournal}



\end{document}